\documentstyle[prl,aps,epsf]{revtex}
\begin{document}
\draft
\twocolumn[\hsize\textwidth\columnwidth\hsize\csname @twocolumnfalse\endcsname

\title{Collective Interaction-Driven Ratchet for Transporting Flux Quanta} 
\author{C.J.~Olson$^{1,4}$, C.~Reichhardt$^{1,4}$, B.~Jank{\' o}$^{2,4}$, 
and Franco Nori$^{3,4}$}
\address{
$^1$ Theoretical and Applied Physics Divisions, Los Alamos National Laboratory, Los Alamos, NM 87545 \\
$^2$ Department of Physics, University of Notre Dame, Indiana 46556 \\
$^3$ Center for Theoretical Physics and Physics Department, University of Michigan, Ann Arbor, 
Michigan 48109-1120 \\
$^4$ Materials Science Division, Argonne National Laboratory, 
Argonne, Illinois 60439 
}
\date{\today}
\maketitle
\begin{abstract}  
We propose and study a novel way to produce a DC transport of 
vortices when applying an AC electrical current to a sample.
Specifically, we study superconductors with a graduated 
random pinning density, which transports interacting vortices 
as a ratchet system. 
We show that a ratchet effect appears as a consequence 
of the long range interactions between the vortices. 
The pinned vortices create an asymmetric periodic flux density profile, 
which results in an asymmetric effective potential for the unpinned
interstitial vortices.
The latter exhibit 
a net longitudinal rectification under an applied 
transverse AC electric current. 
\end{abstract}

\vskip2pc]

\narrowtext

Stochastic transport on asymmetric potentials, the ratchet effect, has 
been recently studied in the context of biology, physics, and applied mathematics.
The central question in these systems is how the random (Brownian) motion 
of a particle is rectified in a spatially asymmetric system.
This can result in a net transport or current of the particle
even in the absence of an 
external DC drive \cite{intro2,fabio3,doering4,feynman5}. 
The ratchet effect is important 
for certain motor proteins and molecular motors. 
In addition, technological applications such as 
new particle separation techniques \cite{Separation} 
and smoothing of atomic surfaces during
electromigration \cite{Lee} also utilize the ratchet effect.
Most studies of ratchets have been conducted for a {\it single\/} 
particle; however, often systems contain many particles, and the 
{\it collective\/} interactions between these particles may significantly 
change the transport properties from the single particle case.
Moreover, most ratchet studies focus on a single particle moving 
on a {\it one}--dimensional (1D) asymmetric potential, as opposed 
to motion in 2D or 3D.

In this work, we propose and study a new type of {\it 2D\/} ratchet 
system which utilizes gradients of point-like disorder, rather than a 
uniformly varying underlying potential. 
In particular, we study fluxons 
in superconductors containing
a periodic arrangement of a graduated density of point defects, 
a geometry motivated by recent experiments \cite{wai10}.
Such defects can be created by either controlled irradiation techniques 
or direct-write electron-beam lithography.
For a sufficiently large externally applied magnetic field, 
the fluxons fill most of the pinning sites and create a
periodic asymmetric, or saw-tooth, 2D flux profile.  A certain 
field-dependent fraction of the vortices do not become pinned 
at individual pinning sites but can move in the interstitial 
regions between pinning sites 
in the presence of an applied AC drive. 
Although the moving interstitial vortices do not directly interact 
with the short-ranged pinning sites, they feel the long-range 
interaction of the vortices trapped at the pinning sites 
and therefore move in an effective asymmetric potential.
The coherence length $\xi$ provides the length scale controlling 
pinning, and this is much smaller than the length scales 
of interactions (given by the penetration depth $\lambda$).
For finite temperature $T$ and for an applied transverse AC 
electric current, we observe a net DC longitudinal transport 
of interstitial fluxons.   Devices built using these ideas 
could be useful for removing unwanted flux in SQUIDs, and for
making devices where flux can be focused via lensing. This 
type of ratchet may also be useful for the transport of 
colloids and charges, in which point defect gradients 
can be constructed.

The ratchet system described here differs significantly from 
other recent proposals for creating ratchets in superconductors.
These range from the use of Josephson junctions in SQUIDs and 
arrays \cite{zapata6}, to the use of a standard 1D-type 
potential-energy ratchet (e.g., \cite{intro2,fabio3,doering4})
to drive fluxons out of superconducting samples \cite{notre7}.
The concept of 2D asymmetric channel walls was proposed 
in Ref.~\cite{michigan8}.  
All of these ratchet proposals rely on single particles interacting with
an external potential to
produce the DC response, whereas in our system collective interactions
are required to produce DC transport.

In order to create ratchet potentials in actual superconducting samples,
an easily controllable method of introducing pinning into the material
is required.  The ratchet geometry described here can be created with
existing experimental techniques.
For example, irradiating the sample with heavy ions produces columnar 
pinning, which is very effective at trapping the vortices and much 
stronger than naturally occurring pinning, except at low temperatures
\cite{Columnar9}.  In a recent experiment \cite{wai10},
controlled irradiation was used to create columnar pinning 
with a sawtooth-shaped, spatially varying density, producing 
a ratchet potential that can induce collective transport 
\cite{wai10} proposed in this research.

We find that the ratchet geometry studied here is effective at 
transporting flux when operated at fields above the first matching 
field, where the density of vortices equals the density of pinning sites.  
For a given driving frequency, the ratcheting effect is optimized for 
narrow ranges in temperature and AC drive amplitude.

\begin{figure}
\centerline{
\epsfxsize=3.5in
\epsfbox{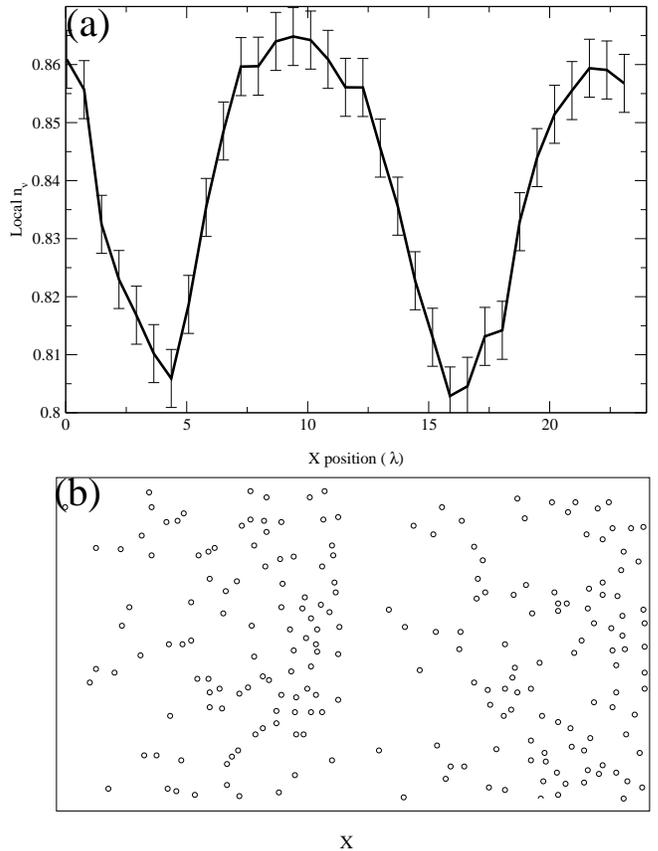}}
\caption{ (a) Local flux density $n_v$ as a function of position X
in a $24\lambda \times 36\lambda$ sample. 
(b) Locations of pinning sites in the sample showing a sawtooth gradient
in pinning density.
The variation in pinning density leads to a variation in the vortex density.
}
\label{fig:pinning}
\end{figure}

{\it Simulation.---\/}
We consider a 2D slice of a system of superconducting vortices interacting
with a pinning background \cite{previouswork10a}.  
The applied magnetic field is ${\bf H}=H{\bf{\hat z}}$, 
and we use periodic boundary conditions in $x$ and $y$ coordinates.
The overdamped equation of motion for a vortex in a bulk superconductor is:
\begin{equation}
{\bf f}^{(i)} = {\bf f}_{vv}^{(i)} + {\bf f}_{L}(t) + 
{\bf f}_{p}^{(i)} + {\bf f}_{T} = {\bf v}^{(i)} \ ,
\end{equation}
where the total force on vortex $i$ due to the repulsion from other vortices  
is given by
${\bf f}_{vv}^{(i)} $
$= \sum_{j=1}^{N_{v}}f_{0} \ K_{1}(|{\bf r}_{i} - {\bf r}_{j}|/
\lambda) \, {\bf {\hat r}}_{ij} .$
Here, 
$K_1$ is a modified Bessel function, 
${\bf r}_{i}$ (${\bf v}_{i}$) is the location (velocity) of the $i$th vortex,
$N_{v}$ is the number of vortices, $f_0=\Phi_{0}^{2}/8\pi^{2}\lambda^{3}$,
and
${\bf {\hat r}}_{ij} =({\bf r}_{i} - {\bf r}_{j})
/|{\bf r}_{i} - {\bf r}_{j}|$. 
We measure 
all lengths in units of the penetration depth $\lambda$.
Most of the results presented here are for systems of size 
$24\lambda \times 18\lambda$ containing $N_v=336$ vortices;
we also considered samples up to $192\lambda \times 36\lambda$ containing
$N_v=5624$ vortices.
The Lorentz driving force from an applied AC current
${\bf J}=J{\bf {\hat y}}\sin{\omega t}$ is modeled as a uniform driving force
${\bf f}_L\sin{2\pi\nu t}$ on the vortices in the $x$-direction.
The pinning sites in the material are assumed to be randomly placed
columnar defects (such as are created by irradiation 
in the experiment of Ref.~\cite{wai10}) 
and are modeled by parabolic traps of radius $r_p=0.2\lambda$.
Each vortex experiences a pinning force of 
${\bf f}_{p}^{(i)} = (f_{p}/r_{p})(|{\bf r}_{i} - {\bf r}_{k}^{(p)}|)
\Theta(r_{p} - |{\bf r}_{i} - {\bf r}_{k}^{(p)}|)
{\bf {\hat r}}^{(p)}_{ik}$,
where ${\bf r}^{(p)}_{k}$ is the location of pin $k$,
$\Theta$ is the Heaviside step function, and
${\bf {\hat r}}^{(p)}_{ik} = ({\bf r}_{i} - 
{\bf r}^{(p)}_{k})/|{\bf r}_{i} - {\bf r}^{(p)}_{k}|$.
We take $f_{p}=2.0f_{0}$.
The samples 
have a sawtooth spatial distribution of pinning site density, repeated
every $12\lambda$ (see Fig.~\ref{fig:pinning}), 
which serves to produce a ratchet potential
in a manner described below.
The forces due to thermal fluctuations, ${\bf f}_T$, are implemented
via Langevin white noise.  Our dimensionless temperature is
$T = k_B T_{\rm actual} / \lambda f_0$.  

{\it Rectification.---\/}When we apply an AC transverse electrical current 
to a sample at finite temperature, we observe a slow net DC longitudinal 
motion of fluxons in the positive $x$ direction, indicating that we 
have succeeded in creating a vortex rectifier or diode.
The individual pins, which interact with the vortices only 
over a {\it short} range, cannot by themselves produce the type
of potential required to generate a ratchet.  Instead, it is
pinned vortices, which interact with unpinned vortices over a 
much {\it longer} range, that provide the properly shaped potential.  
For the strong pinning strengths considered here, a vortex 
that is trapped by a pinning site never depins afterwards.  
Since we focus in the regime above the first matching field, 
there are more vortices than pinning sites.  Thus, not all 
of the vortices can be trapped.  The mobile interstitial 
vortices are the ones that participate in the 

\hspace{-13pt}
rectified 2D motion.  
They interact with the gradient in vortex density established by the
gradient in pinning density.
The local vortex density $n_v(x)$ integrated along the 
$y$-direction is shown in Fig.~\ref{fig:pinning}(a).  
Fig.~\ref{fig:pinning}(b) shows 
the top view, along $z$, of the $x$--$y$ cross-section of the sample, 
and the location of the pinning sites (open circles).  
A flux profile similar to that in Fig.~\ref{fig:pinning}(a)
has been observed experimentally through magneto-optical imaging of
an irradiated sample containing pins arranged as in 
Fig.~\ref{fig:pinning}(b) \cite{UWelp}.
This structure, made of columnar pins, provides a ratchet that 
works based on the collective interactions of the movable objects. 
This type of structure would not function for a single vortex, but 
requires many interacting vortices to be present in order for it to work.

\begin{figure}
\centerline{
\epsfxsize=3.5in
\epsfbox{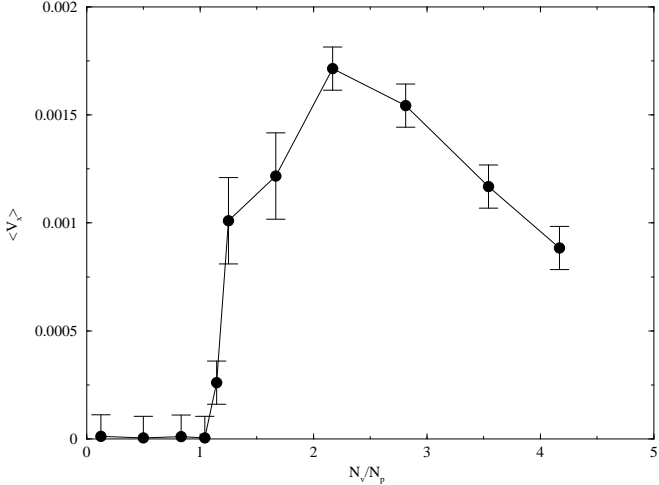}}
\caption{
The dependence of DC voltage  $\langle v \rangle$ on the ratio of
vortices to pins $N_v/N_p$ in the sample for temperature $T=0.045$,
driving force $f_L=0.1f_0$ and frequency $\nu=0.003125$. 
An optimal filling fraction appears near $N_v/N_p=2$.  Below
$N_v/N_p=1$ no DC signal appears because all of the vortices are
trapped in pinning sites.
}
\label{fig:field}
\end{figure}

{\it Field.---}To further illustrate the collective nature of the 
rectification process, 
in Fig.~\ref{fig:field} we show the rectified net fluxon velocity 
$ \langle v \rangle = \sum_i^{N_v} {\bf {\hat x}} \cdot v_i / N_v$
obtained as the vortex density in the sample is varied, for a
fixed temperature of $T=0.045$.  
When there are fewer vortices than pins, $N_v/N_p<1$, each vortex 
is trapped by a pin during the initial annealing period.  When 
a driving force of amplitude $f_L=0.1f_0$ and frequency $\nu=0.003125$ 
is applied, the vortices are 
unable to escape
from the 

\hspace{-13pt}
pinning sites and remain stationary.  Thus, we see no ratchet
signal when there are no unpinned vortices: $\langle v \rangle =0$
for $N_v/N_p<1$.  Once all of the pins have been filled, additional
vortices remain in the interstitial region between pins.  The motion of
these interstitial vortices is influenced by the pinned vortices, but
not directly by the short-ranged pins.  As soon as interstitial vortices
appear, a ratchet signal is obtained, as clearly indicated by the 
abrupt increase in $\langle v \rangle$ above $N_v/N_p=1$ 
in Fig.~\ref{fig:field} \cite{Note}.
Initially, the magnitude of the ratchet signal continues to increase as 
the number of interstitial vortices increases, but at higher applied 
magnetic fields the strong vortex-vortex interactions begin to inhibit 
the ratcheting motion.  In Fig.~\ref{fig:field} $\langle v \rangle$ 
begins to decrease slowly above $N_v/N_p=2$.

\begin{figure}
\centerline{
\epsfxsize=3.5in
\epsfbox{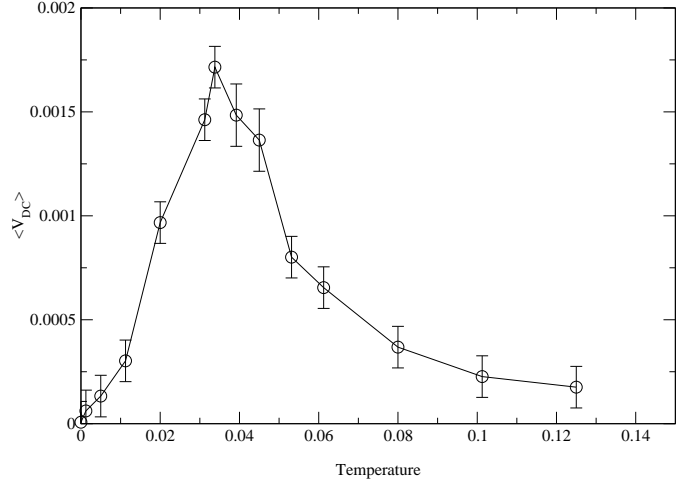}}
\caption{
The dependence of DC voltage  $\langle v \rangle$ on temperature $T$ for
a sample with driving force $f_L=0.1f_0$ and frequency $\nu=0.003125$. 
An optimal temperature appears near $T=0.035$.
}
\label{fig:tempe}
\end{figure}

{\it Temperature.---}We explore the properties of the vortex
diode under varying conditions by measuring $V_x$.
Initially, we apply a driving force of amplitude $f_L = 0.1f_0$
and frequency $\nu=0.003125$, and vary the temperature from 
$T=0$ to $T=0.125$.  The resulting $\langle v \rangle$ 
is shown in Fig.~\ref{fig:tempe}.
These simulations clearly indicate an optimal or ``resonant'' 
temperature regime in which the DC fluxon velocity is maximized 
by the fluxon pump or diode.  This optimal temperature regime 
can be explained as a trade-off between allowing the fluxons 
to fully explore the ratchet geometry, and washing out the 
driving force or pinning at high temperatures.

\begin{figure}
\centerline{
\epsfxsize=3.5in
\epsfbox{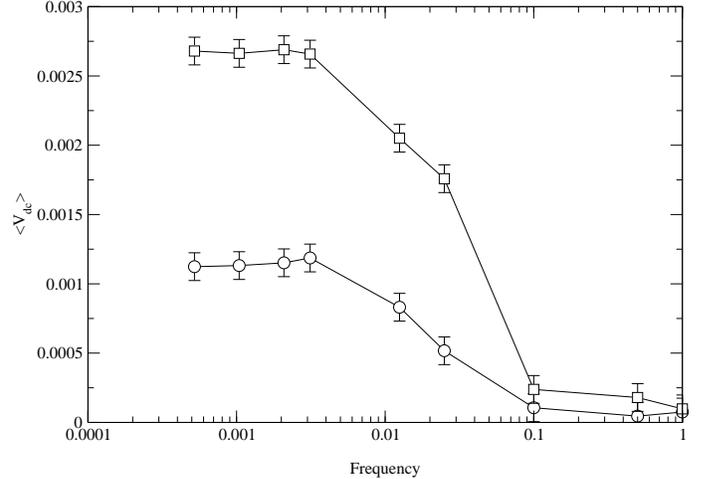}}
\caption{
The dependence of DC voltage $ \langle v \rangle $ on frequency $\nu$
for a sample with $T=0.045$ at two different amplitudes:
circles, $f_L=0.1f_0$; squares, $f_L=0.2f_0$.  The rectification is
reduced at high frequencies and saturates at low frequencies.
}
\label{fig:freq}
\end{figure}

{\it Frequency.---}We show the fluxon velocity $<v>$ at 
varying frequencies for a sample at $T=0.045$ at two 
different driving forces of $f_L=0.1f_0$ and $f_L=0.2f_0$
in Fig.~\ref{fig:freq}.  
The ratcheting effect dies away at high frequencies because 
the vortices do not have enough time to  explore the 
ratchet potential. We also observe a 
saturation of the 

\hspace{-13pt}
voltage response at low frequencies as we reach the DC limit.
This is because in the low-frequency limit, the interstitial 
vortices can repeatedly sample the asymmetry of the underlying ratchet
potential, and the difference in the transport current in the two
directions is fully exploited.

\begin{figure}
\centerline{
\epsfxsize=3.5in
\epsfbox{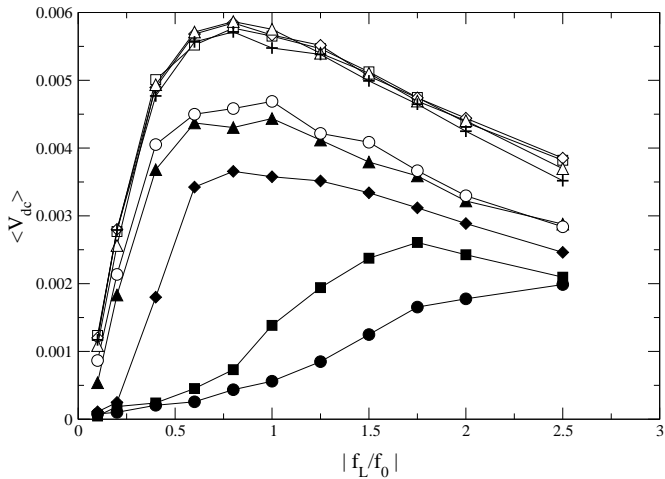}}
\caption{
Rectified average fluxon velocity  
$\langle v \rangle$, 
which can be measured as a voltage, 
versus the amplitude of the AC driving force.
The frequency of the AC signal is: 
filled circles, $\nu=1$; filled squares, $\nu=0.5$; 
filled diamonds, $\nu=0.1$; filled triangles, $\nu=0.025$; 
open circles, $\nu=0.0125$; open squares, $\nu=0.003125$; 
open diamonds, $\nu=0.0021$; open triangles, $\nu=0.00104$; 
pluses, $\nu=0.000521$.
}
\label{fig:ampl}
\end{figure}

{\it Amplitude.---}We find that there is an optimal 
driving amplitude, as shown in Fig.~\ref{fig:ampl} for samples at
$T=0.045$ driven with frequencies ranging from $\nu=0.000521$ to $\nu=1$.
The optimal amplitude shifts to higher driving forces as the 
driving frequency is increased, since the vortex can explore the
same portion of the pinning in a shorter time interval if the 
driving force is increased.

{\it Conclusion.---\/}
Using a new type of ratchet system that fundamentally depends 
on the collective interactions of the movable objects, 
we observe a DC vortex rectification starting with an input 
AC electrical current.  Our proposal employs superconductors with 
periodic, graduated random pinning density of columnar defects.  
This ratchet differs from the majority of the ones proposed previously 
on several key points: (a) it is fundamentally a 2D ratchet, as opposed 
to a 1D one, and more important (b) it requires collective interactions, 
as opposed to the mostly one-particle ratchets studied so far.
In our system, we show that the asymmetric potential is created 
when vortices are trapped by the pinning sites and interact with 
unpinned vortices.  The long range interactions of the vortices 
leads to the formation of a periodic asymmetric potential caused
by the gradient in vortex density.  
Mobile interstitial vortices experience this collectively-produced 
potential, rather than interacting directly with the pinning sites 
in a single-particle manner.  We show that there is an optimal
field, temperature, and frequency for the operation of such devices.
This ratchet can be created experimentally through controlled 
irradiation techniques and via electron-beam lithography.
The use of controlled random defects in the ratchet geometry proposed here
should make it possible to extend the rectification devices shown
here to other systems,
including colloids and other collections of charged particles.

We gratefully acknowledge G. Crabtree, W. Kwok, F. Marchesoni,
V. Vlasko-Vlasov, and U. Welp for very useful discussions.
This work was partially supported by 
DOE Office of Science under contract 
No. W-31-109-ENG-38, 
%CLC and CULAR (LANL/UC),
the NSF DMR-9985978, 
the Michigan Center for Theoretical Physics (MCTP), 
and the Center for the Study of Complex Systems 
at the University of Michigan.  This is preprint MCTP-00-19.


\begin{references}
\bibitem{Note} Due to the gradient in pinning density, the field at which 
$N_v=N_p$ is {\it lower} than the field at which the vortex density equals 
the maximum pinning density in the sample.
\end{references}
\end{document}